\documentclass[aps, prb, twocolumn,floatfix,superscriptaddress,longbibliography, nofootinbib, sortedaddress]{revtex4-2}

\usepackage{graphicx, color, soul}
\usepackage{amsmath, amsfonts, amssymb, mathrsfs, dsfont, mathtools}

\usepackage[usenames,dvipsnames, pdftex, table]{xcolor}
\usepackage[normalem]{ulem} % needed for \scrap-command
\usepackage[mathscr]{eucal}
\usepackage{bm}
\usepackage{tabularx, booktabs}
\usepackage{multirow}
\usepackage{physics}

\usepackage[colorlinks, breaklinks, 
            linkcolor=Blue,
            citecolor=Blue,
            urlcolor=Blue]{hyperref}
            
\usepackage[all]{hypcap} 
\usepackage{enumitem}

\newcommand{\Le}{\left}
\newcommand{\Ri}{\right}
\newcommand{\nn}{\nonumber}

\newcommand{\dg}{\dagger}
\newcommand{\ra}{\rangle}
\newcommand{\la}{\langle}
\newcommand{\eq}[1]{\begin{align}#1\end{align}}

\newcommand{\msr}{\mathscr}
\newcommand{\hc}{h_\text{c}}
\newcommand{\br}{\bm{r}}

\date{\today}

\begin{document}
% \title{Delocalization transition of disordered Hopf insulator}
\title{Disorder-induced delocalization and reentrance in a Chern-Hopf insulator}

\author{Soumya Bera}
\affiliation{Department of Physics, Indian Institute of Technology Bombay, Mumbai 400076, India}

\author{Ivan Dutta}
\affiliation{National Institute of Science Education and Research, Jatni, 752050, India}
\affiliation{Homi Bhabha National Institute, Training School Complex, Anushakti Nagar, Mumbai 400094, India}

\author{Roderich Moessner}
\affiliation{Max-Planck Institute for the Physics of Complex System, Dresden, Germany}

\author{Kush Saha}
\affiliation{National Institute of Science Education and Research, Jatni, 752050, India}
\affiliation{Homi Bhabha National Institute, Training School Complex, Anushakti Nagar, Mumbai 400094, India}

\begin{abstract}
The Chern-Hopf insulator is an unconventional three-dimensional topological insulator with a bulk gap and gapless boundary states without protection from global discrete symmetries. 
This study investigates its fate in the presence of disorder. 
We find it stable up to moderate disorder by analyzing the surface states and the zero energy bulk density of states using large-scale numerical simulation and the self-consistent Born approximation. 
The disordered Chern-Hopf insulator shows reentrant behavior: the disorder initially enhances the topological phase before driving it across an insulator-diffusive metal transition.
We examine the associated critical exponents via finite-size scaling of the bulk density of states, participation entropy, and two-terminal conductance. 
We estimate the correlation length exponent $\nu\simeq 1.0(1)$, consistent with the {\it clean} two-dimensional Chern universality and distinct from the integer quantum Hall exponent. %\SB{Note: $\nu=1$ is the clean exponent found either using the Chern number scaling or Chern marker %scaling.. in the presence of disorder it flows to the IQHE exponent.}
\end{abstract}
\maketitle

\section{Introduction}

%{\color{blue}
It is generally understood that for disorder-driven quantum phase transitions, i.e., Anderson transitions, the symmetry of the underlying Hamiltonian and its spatial dimensions solely determine the universality class and the corresponding critical exponents~\cite{EversRMP08}.  
Whether or not introducing band topology will preserve the universality class of the transition depends on the underlying symmetries of the disordered potential and has been a subject of intense research in the last couple of decades, e.g., see Refs.~\cite{TomiPRL02, NagaosaPRL07, MirlinIJMP10, nyu_NJP2010,goswami_PRL2011,fukane_PRL2012, BitanPRL17}. 
While non-spatial symmetries of standard topological insulators are expected to protect the surface states~\cite{Prodan_JPA2011, onada_PRL2007}, the disorder can disrupt them~\cite{nomura_PRL2007,he_PRL2011,lu_PRL2011}. Interestingly, these states remain stable under weak to moderate disorder as long as symmetry and the bulk gap are preserved~\cite{vojtadisorderPRB2012}. Adding disorder can even enhance the topological phase by effectively renormalizing system parameters from an initially trivial parameter regime~\cite{GuoPRL10, groth_PRL2009,shun_PRL2009,Prodan_2011,kuramoto_jsps2011,prodhan_PRB2011,jian_PRB2011,leung_PRB2012,pradhan_PRB2014,sriluckshmy_PRB2018,yang_PRB2021,hu_PRB2021}.
Eventually, increasing disorder drives the system into a diffusive metallic regime and induces standard Anderson localization, transitioning the system from a diffusive metal to an insulating phase~\cite{RyuPRB12,song_PRB2012,xu_PRB2012}.
%in three spatial dimensions
% The interplay between disorder and topological states has been a significant research focus for over a decade~\cite{}.

% Understanding the robustness and stability of bulk and boundary states in topological materials under quenched disorder is crucial for developing quantum devices that exploit their unique transport properties arising from topological protection. 
%
%
%
%For instance, symmetry-respecting disorder %protects boundary states
%

The situation becomes intriguing with non-conventional topological insulators outside the standard 10-fold classification scheme~\cite{altland_PRB1997}.
The stability of these insulators, which are protected by generic crystalline symmetries~\cite{mong_PRB2010,hughes_PRB2011,fu_PRL2011,SlagerNatPhy2013,liu_PRB2014,fang_PRB2015,PhysRevX.7.041069} -- such as the axion insulators~\cite{xu_PRL2019,song_PRL2021, LiAxionPRL21} and higher-order topological insulators~\cite{doi:10.1126/science.aah6442,benacazar_PRB2017,doi:10.1126/sciadv.aat0346,ezawa_PRL2018,araki_PRB2019,wang_PRR2020,li_PRL2020,adhip_PRR2020,szabo_PRR2020} -- in the presence of the disorder have received considerable attention in recent years. 
For instance, disorder can induce a bulk quantum phase transition in an inversion-symmetric three-dimensional~(3D) axion insulator, characterized by a localization exponent $\nu \simeq 1.42(10)$~\cite{song_PRL2021}. In contrast, the surface of such an insulator exhibits a phase transition between the axion insulating phase and the Anderson insulating phase driven by magnetic disorder, with the critical point showing 2D quantum Hall criticality characterized by a critical exponent $\nu \simeq 2.6(2)$~\cite{LiAxionPRL21}. 
%\KS{In the similar spirit, semimetals under Euler class has been shown to exhibit distinct critical behavior in the presence of disorder as compared to graphene and 3D Weyl semimetals\cite{robert_arXiv}}. 
%

\begin{figure}[!tb]
    \centering
    \includegraphics[width=1\columnwidth]{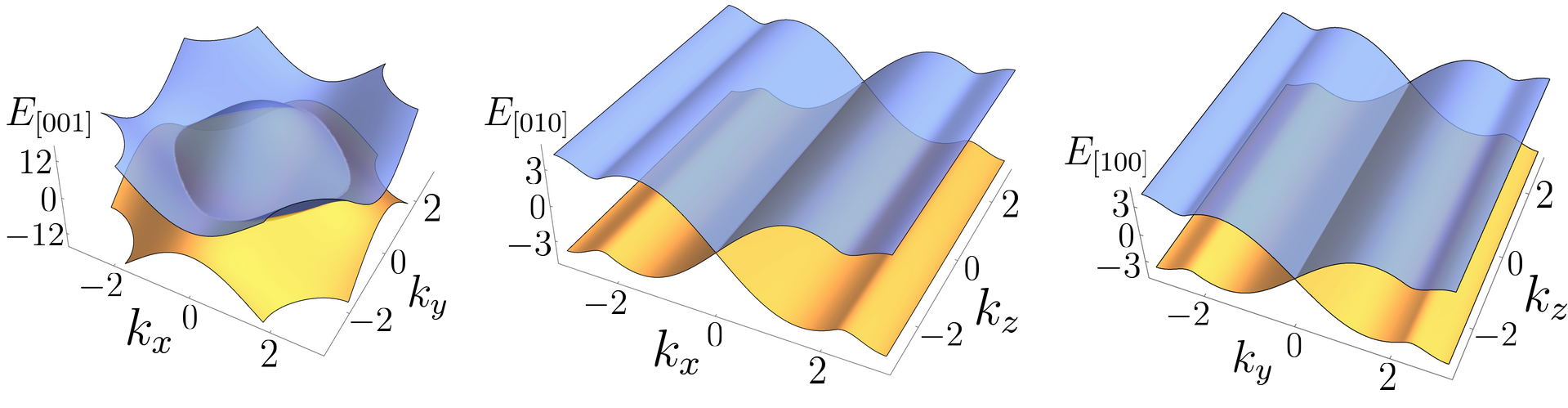}
        \caption{Surface states dispersion along the $(001)$, $(010)$, and $(100)$ directions: a Dirac nodal ring in the $(001)$ plane, and a Dirac line at zero energy in both $(010)$ and $(100)$ planes, differing from typical Dirac nodes on conventional topological insulator surfaces. We use periodic boundary conditions in the directions perpendicular to the surface under consideration. The parameters in Eq.~(\ref{eq:ham2}) are taken to be $t_2{=}-2$, $h{=}-4$, and $t{=}5$.} 
    \label{fig:surfacemodes}
\end{figure}

Another class of unconventional 3D topological insulators\footnote{We note that Hopf insulators are sometimes considered a subclass of axion insulators~\cite{PhysRevB.103.014417}. However, the axion insulators mentioned earlier, protected by inversion symmetry, differ from Hopf insulators, characterized by a combination of time-reversal and reflection symmetries.}, which lack topological protection from conventional discrete symmetries, is the Hopf insulators~\cite{moore_PRL2008,deng_PRB2013,yan_PRL2017,liu_PRB2017,alex_PRB2021,zhu_PRB2021,yue_PRB2023,lapierre_PRR2021,nelson_PRB2022,goswami_arXiv,jankowski2024non} and are characterized by a non-zero Hopf index~\cite{moore_PRL2008,moorePRB2010,PhysRevB.102.115135,PhysRevLett.125.053601,PhysRevResearch.1.022003, PhysRevB.109.155131}.
They are described by a minimal two-band model, $\msr{H}(\bm{k}) = \bm{d}(\bm{k}) \cdot \bm{\sigma}$, where $\bm{\sigma}{=}\{\sigma_x, \sigma_y, \sigma_z\}$ represents a set of Pauli matrices, and $\bm{d}{=}\{d_x, d_y, d_z\}$ encodes a bulk gap that prevents a continuous deformation of the Hamiltonian into a trivial $\bm{k}$-independent Hamiltonian.
From the homotopy perspective, the momentum $\bm{k}$ belongs to the three torus $T^3$, representing the Brillouin zone. 
The unit vector $\bm{d}(\bm{k})/|\bm{d}(\bm{k})|$ identifies a point in the two-sphere $S^2$; thus, the Hamiltonian can be considered a map from $T^3$ to $S^2$. The three 2D planes $xy$, $yz$, and $xz$ of $T^3$ can have non-zero Chern number $C_\alpha$, where $\alpha \in (x,y,z)$. For $C_{\alpha}=0$, the Hopf insulators are constructed by an intermediate map: a point on $T^3$ to a point in three-sphere $S^3$, then to a point in $S^2$ via a Hopf map. This non-trivial mapping results in an integer-valued Hopf invariant.
Conversely, when $C_\alpha \neq 0$, the invariant takes values in the finite group $Z_{2. \text{gcd}(C_{x},C_{y},C_{z})}$, where $\text{gcd}$ denotes the greatest common divisor. These insulators are termed Chern-Hopf insulators (CHI).
We focus on a scenario where any two-dimensional slice of the 3D Brillouin zone carries a finite Chern number~\cite{Kennedy_PRB2016} and features non-trivial gapless boundary states, such as a nodal ring (see Fig.~\ref{fig:surfacemodes}). We aim to understand the effect of disorder on the CHI phase and the corresponding boundary states, which lack protection from non-crystalline symmetry.

%%%%%%%%%%%%%%%%%%%%
%%%%%%%%%%%%%%%%%%%%

We observe that the boundary states remain stable under weak to moderate disorder. However, as the disorder increases, the CHI phase transitions into a diffusive metal phase, qualitatively characterized by the closing of the bulk gap and the disappearance of the boundary states. Additionally, numerical simulations show that disorder enhances the topological phase, a result we confirm using the self-consistent Born approximation (SCBA).
We further investigate the critical properties of this transition through a scaling analysis of several observables, including the density of states $\rho(E)$, participation entropy $\msr{S}_1$, and two-terminal conductance $g$. Remarkably, we observe that the correlation length exponent closely matches the universality class of the two-dimensional clean Chern insulator with $\nu \simeq 1.0(1)$. Finally, we discuss the potential material realization of the Chern-Hopf model.

\section{Model}
%%%%%%%%%%%%%%%%%%%%
A generic 3D tight-binding Hamiltonian, defined on a cubic lattice in which each lattice site contains two orbital states $|A\rangle$, $|B\rangle$, with nearest-neighbor (NN) and next-nearest neighbor hopping (NNN) can be written as~\cite{Dutta_2024}
%\rim{is notation at lower limit of sum as intended?} \KS{sum over $\alpha\beta$ not required as $\gamma$ encodes sublattice hoppings}
\begin{align}
H_\text{3D} &=\sum_{\br,\mu,\nu}| \br+e_{\mu}\rangle\,\gamma^{\mu\nu}\,
\langle \br+e_{\nu}|,
%\sum_{i\sigma}\epsilon_{\sigma}c_{i\sigma}^{\dagger}c_{i\sigma}+ \sideset{} {_{x} }\sum_{<ij>,\sigma,\sigma'}t_{\sigma\sigma'}^x c_{i\sigma}^{\dagger} c_{j\sigma'} \nonumber \\
%& + \sideset{} {_{y} }\sum_{<<ij>>\sigma,\sigma'}t_{\sigma\sigma'}^y c_{i\sigma}^{\dagger} c_{j\sigma'} 
\label{eq:Ham0}
\end{align}
where $\br$ denotes lattice site, $e_0=(0,0,0)$ is the  null vector and $e_{\mu (\nu)}$ is the vector encoding the six-fold coordination via $e_{1,....6}=\{x,y,z, xy,xz,yz\}$, respectively. The coordinates, for example, $x$, $xy$ represent vectors  $(1,0,0),(1,0,1)$, respectively, $\gamma^{\mu\nu}$ are hopping matrices involving sublattices. With this, the minimal CHI model  can be constructed with the following matrices:
\begin{align}
&\gamma^{0,0}=h\,\sigma_z;\gamma^{1,0}=t\,\sigma_z;\gamma^{2,0}=t\,\sigma_z; \nonumber\\
&\gamma^{5,0}=-\frac{t_2}{2}(\sigma_x+i\sigma_y); \gamma^{1,3}=\frac{t_2}{2}(\sigma_x-i\sigma_y);\nonumber\\
&\gamma^{6,0}=\frac{t_2}{2}(\sigma_y-i\sigma_x); \gamma^{3,2}=\frac{t_2}{2}(i\sigma_x-\sigma_y),
\label{eq:ham2}
\end{align}
where $\sigma_i$'s are the Pauli matrices and $h,\,t,\,t_2$ are the model parameters. This is equivalent to constructing a 2D Chern model in the $x-y$ plane and extending it in the $z$-direction to construct the 3D model. 
The corresponding momentum space Hamiltonian takes the following form,  $H({\bm k})=d({\bm k})\cdot\sigma$, where
\eq{
d_x(\bm{k}) &= 2 t_2 (\sin k_x \sin k_z +\cos k_z \sin k_y), \nn \\
d_y(\bm{k}) &= 2 t_2 (-\sin k_y\sin k_z +\cos k_z \sin k_x), \nn \\
d_z(\bm{k}) &= h+2t(\cos k_x+\cos k_y). 
\label{eq:continuum}}
We note that a minimum tight-binding model of a Hopf insulator without any non-zero Chern number requires NNNN hopping~\cite{Dutta_2024}.

The CHI Hamiltonian in Eq.~\ref{eq:Ham0} exhibits two topological invariants: Chern number ($C_z=1$) in the $xy$ plane and a 3D Hopf invariant ($h_f=1$). For $-4t<h<0$, we obtain a gapped insulating phase with $C_z=1$ and for $0<h<4t$, we obtain $C_z=-1$. In both cases, the Hopf invariants are found to have integer values with $|h_f|=1$. The bulk gap closes at $h=|4t|$, and for $h>|4t|$, we obtain a trivial insulating phase with both $C_z$ and $h_f$ to be zero. 

To find the surface states for different planes as shown in Fig.~(\ref{fig:surfacemodes}), we numerically solve the tight-binding Hamiltonian in Eq.~(\ref{eq:Ham0}) with open boundaries along $z$, $y$ and $x$, respectively. In certain cases, the effective Hamiltonian for the surface states can be obtained numerically, as shown in Appendix~\ref{app:surfaceHam}.  The surface states of $(001)$ turn out to be a nodal ring, while the surface states for $(010)$ and $(100)$ are line Dirac nodes. 
%Interestingly, the $d_z$ component of the Hamiltonian turns out to be manifested as the effective surface Hamiltonian for the $(001)$ plane. Likewise, the $d_y$ and $d_x$ components of the Hamiltonian are reflected in the boundaries of $(010)$ and $(100)$ planes, respectively. 

To address the effect of disorder, we add onsite local magnetic disorder\footnote{The critical properties of the Hopf-insulator to diffusive metal transition are unaffected by local charge impurity disorder instead of magnetic disorder. The corresponding data is shown in App.~\ref{app:onsite}.} to Eq.~(\ref{eq:Ham0}), in the form $H_{\rm dis}=\sum_{\br}  | \br\rangle_{\alpha} U_{\alpha\alpha} ~_\alpha\langle \br |$, where $U_{\alpha\alpha}=U(\br)\sigma_z$. 
%\rim{indices look weird here, I guess $\alpha\beta$ are implicit in the matrix form of $\sigma$?} %\KS{Agree: It will be simply $U$} 
We consider the disorder potential $U(\br)$ distributed uniformly over the interval $[-W, W]$. 
In the continuum description, we take $\langle U(\br)\rangle=0$ and, $\langle U(\br)U(\br')\rangle=\frac{W^2}{3}\delta(\br-\br')$, where $\delta(\br)$ is the Dirac delta function. {In all numerical simulations we choose  $t=1.0$, and $t_2=1.0$ unless mentioned otherwise.}

%%%%%%%%%%%%%%%%%%%%
\begin{figure}[!t]
    \centering
    \includegraphics[width=1\columnwidth]{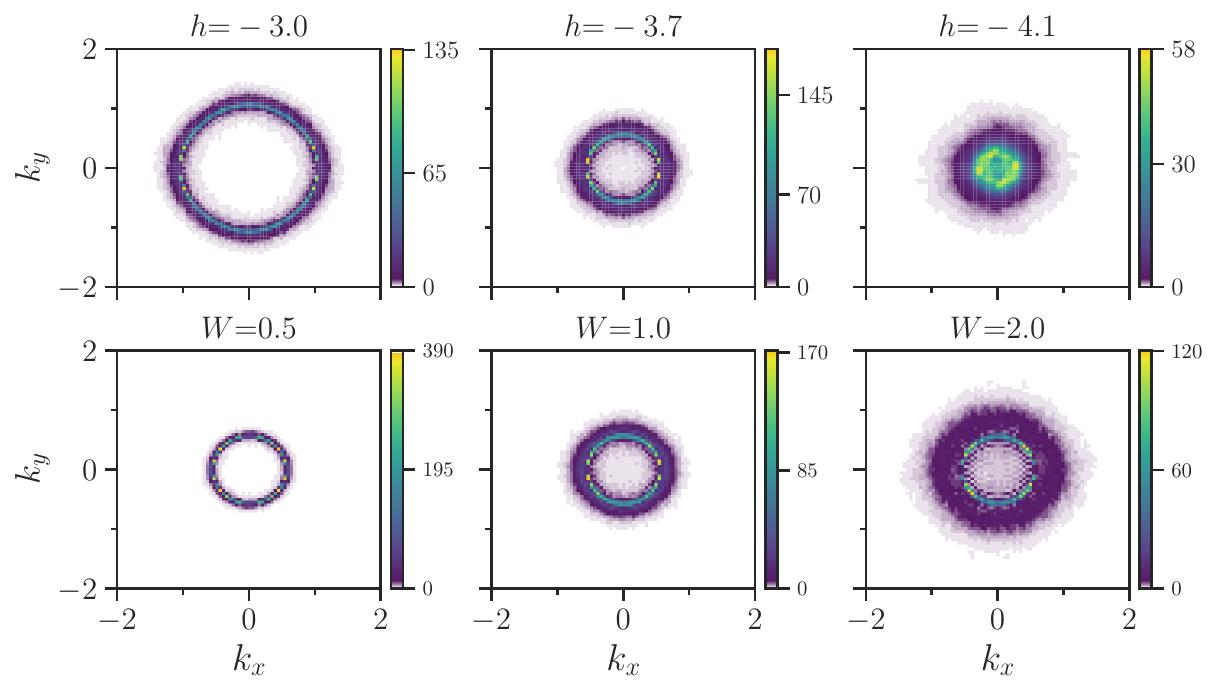}
        \caption{Evolution of the surface state along the (001) direction as seen in momentum resolved surface spectral function, $\msr{A}(k, E)$~\eqref{eq:spectral} at  $E=0$ (color bar represents the value in arbitrary units). Top row:  for different $h$-parameter as defined in Eq.~\eqref{eq:ham2}. Bottom row: for different disorder strengths $W$. The delocalization transition is signified by the decrease in the radius of the nodal ring in the upper panel and by the decrease in the spectral weight of the surface mode in the lower panel. The system size is \{$L_x, L_y, L_z\}=\{128,  128,  64\}$.
        % associated with the destruction  \rim{it is not clear what 'destruction' means/how it is visible in this figures}
        } 
    \label{fig:dis_surface}
\end{figure}
%%%%%%%%%%%%%%%%%%%%

%%%%%%%%%%%%%%%%%%%%
\begin{figure*}[!th]
    \centering
    \includegraphics[width=1\textwidth]{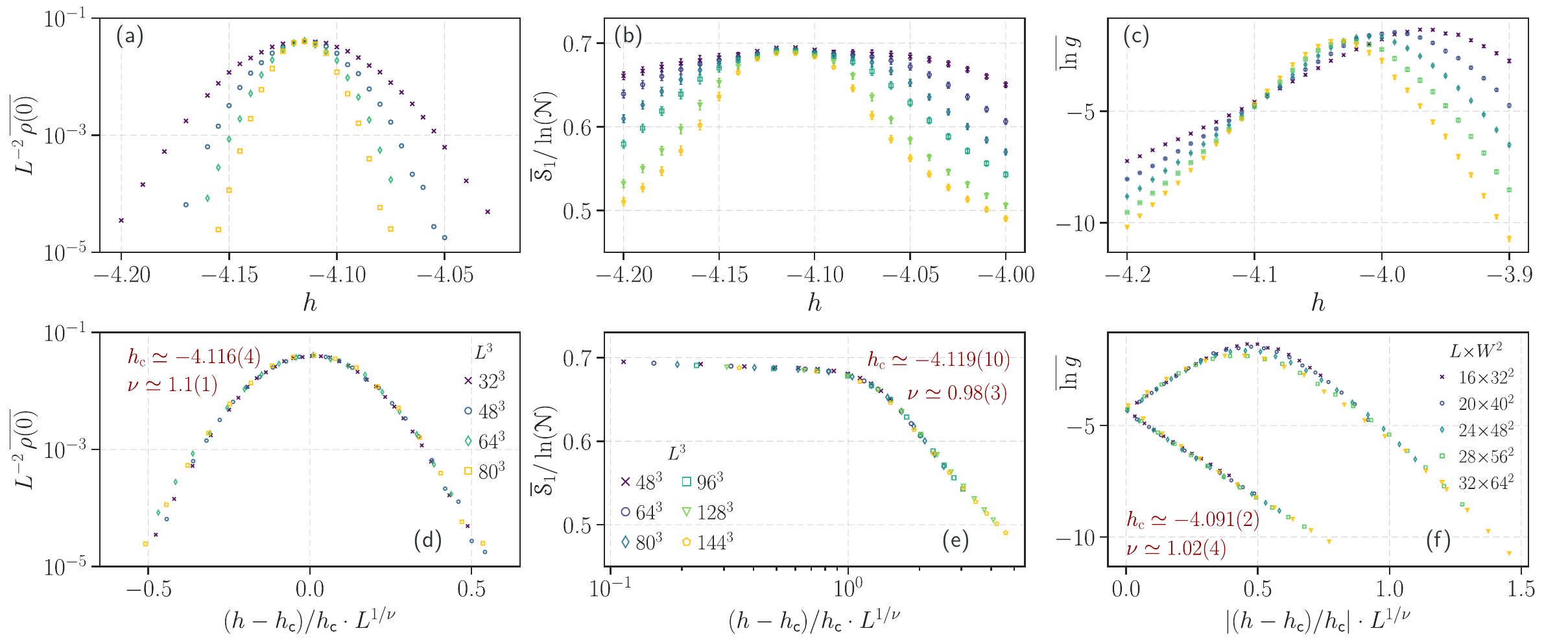}
        \caption{(a)~The zero energy DOS $\overline{\rho}(\varepsilon=0)$ for cubes of linear length of $L=32,48, 64, 80$ across the  transition driven by $h$. (b)~Renormalized participation entropy $\overline{\msr{S}_1}/\ln(\msr{D})$ across the same transition for different $L$. (c)~Two terminal conductance $\overline{\ln (g)}$ at $\varepsilon=0$ for samples of length $L$ along the transport direction, and $W$ the width of the perpendicular direction (see panel (f)). (d-f)~Corresponding scaling collapses with $\hc\simeq -4.11$, and $\nu\simeq 1.0$, consistent within the least-square fitting errors (mentioned in the plots) for all the observables. }
        %\rim{right panels: not clear to me how rescaling x-axis makes function multiply-valued--missing modulus signs?}} 
    \label{fig:scaling}
\end{figure*}
%%%%%%%%%%%%%%%%%%%%

%%%%%%%%%%%%%%%%%%%%
\section{Results and discussion}
%%%%%%%%%%%%%%%%%%%%
%
\paragraph*{Stability of the surface mode:} 
Probing the stability of surface states against disorder provides insights into their topological protection, as discussed previously for topological insulators~\cite{vojtadisorderPRB2012} and, more recently, for Weyl surface states~\cite{SlagerPRB17}. 
We start discussing the evolution of the surface mode in the (001) direction of the Hopf insulator~\eqref{eq:Ham0} in the presence of onsite disorder. 
The momentum-resolved boundary spectral function is defined as, 
\eq{
\msr{A}(\bm{k}, E) =  \sum_{j=1}^{\msr{D}} \sum_{\alpha=1}^2  | \psi^{\text{s}}_j(k_x, k_y, \alpha)|^2 \delta (E - E_j), 
\label{eq:spectral}
}
where $j$, $\alpha$ denote the spatial index and the orbital degrees, respectively; 
$\psi^{\text{s}}_j(\vec{k}, \alpha)$ denotes the $E=0$ momentum resolved surface wavefunction at the (001) boundary. 
We approximate the delta function in Eq.~\eqref{eq:spectral} with a box function by considering only four surface states close to zero energy, calculated using standard diagonalization techniques. 

The upper panel of Figure~\ref{fig:dis_surface} illustrates the spectral function across various $h$ values with a fixed disorder of $W=1.1$. 
The nodal ring at $E=0$ broadens due to disorder, and its radius decreases as $h$ nears the gap-closing threshold. 
Near the critical point, $h_c\simeq-4.1$, the ring structure ceases to exist, signaling a transition to a metallic state without the topological surface state. 
%
% \rim{how does disappearnce of ring structure imply absence of edge states?}

In the lower panel of Fig.~\ref{fig:dis_surface}, the same surface mode is shown for a fixed value $h=-3.7$ with increasing disorder. 
The state survives at weak disorder; however, with spectral broadening, a typical response to the disorder, {as also observed in} 3D topological insulators~\cite{vojtadisorderPRB2012}. 
As disorder increases beyond $W\gtrsim 1$, the ring expands progressively until it disappears, qualitatively coinciding with the closure of the bulk gap, as seen in Fig.~\ref{fig:phasediag}. 
In App.~\ref{app:xsurface}, the evolution of the topological surface state in the (100)-direction is shown, corroborating the similar physics observed in the (001) direction.

\paragraph*{Density of states~(DOS):}
The disorder averaged bulk DOS $\overline{\rho}(\varepsilon=0)$ acts as an order parameter for the transition; here, overline denotes the averaging over different disorder configurations.  %\rim{what's configuration averaging?}. 
The transition to diffusive metal is signaled by a finite density of state at $\varepsilon=0$ as shown in Fig.~\ref{fig:phasediag}.
The raw data calculated by diagonalizing the Eq.~\eqref{eq:Ham0} is shown in Fig.~\ref{fig:scaling}(a) for different system sizes $L$ across the transition driven by the field $h$.  
In the metallic phase, the normalized density of state $\tilde{\rho}(0){=}L^{-2} \overline{\rho}(0)$ is finite, while in the localized phase, it vanishes with increasing system size. 
%\SB{Figure out the $L^{-2}$ dependence!}\KS{The reason is %attributed to the effective two-dimensional nature of the spectrum}
%
Generically, the following scaling form  is usually used to perform the finite size scaling analysis, 
%\rim{what is $\phi$?}
\eq{
\tilde{\rho}(0) = \sum_{j=0}^{N_\text{R}} a_j x^j + \mathrm{b} L^{-y} + \mathrm{c} \, \mathrm{b} L^{-y} \, x, 
\label{eq:scaling}
}
where $x=(h-\hc)/\hc \cdot L^{1/\nu}$, the $h_c$ is the critical field, and $\nu$ is the leading scaling exponent, and the $N_\text{R}$ is the expansion order, which usually chosen to be $N_\text{R}=2$, to reduce the number of fitting parameters $a_j, \mathrm{b}, \mathrm{c}$\footnote{In the scaling ansatz, we ignore the irrelevant scaling variables; the justification comes from the observation that in standard 3D disordered models, the irrelevant correction to scaling is small, i.e., $y$ is large~\cite{SlevinPRL99}. Nonetheless, we checked (data is not shown) and found that it does not change the reported critical exponent.}.
We use least-square fitting to determine the coefficients $a_j$, and the critical parameters $h_\text{c}, \, \nu$. 
The finite size collapse of $\rho(0)$ is shown in Fig.~\ref{fig:scaling}(d) with a critical exponent $\nu\simeq 1.1(1)$, and $\hc \simeq -4.116 (4)$. 
%

%%%%%%%%%%%%%%%%%%%%%%%%%%%%%%%%%%%55
\paragraph*{Participation entropy:}
%%%%%%%%%%%%%%%%%%%%%%%%%%%%%%%%%%%55
Across the transition, we probe the nature of the state through the participation entropy~\cite{JiaPRB08}, defined as 
$$
\msr{S}_1 = - \sum_{\br} \sum_\alpha |\psi_\alpha(\br)|^2 \ln |\psi_\alpha(\br)|^2, 
$$
where $\psi_\alpha(\br)$ is the bulk wavefunction at $E=0$ calculated using shift-invert diagonalization. 
In the insulating phase,  $\msr{S}_1$ is constant for finite $L$, while in the metal phase $\msr{S}_1= \ln \msr{D}$, where $\msr{D}$ is the Hilbert space dimension. 
Figure~\ref{fig:scaling}(b) shows the renormalized $\overline{\msr{S}}_1/\ln \msr{D}$, for different system sizes, across the transition. 
At the critical point, the participation entropy is finite, signaling a metallic point, while in the localized phase, increasing $L$ makes it vanish trivially. 
The corresponding collapse, Fig.~\ref{fig:scaling}(e), of $\overline{\msr{S}}_1/\ln \msr{D}$, obtained using the same scaling function~\eqref{eq:scaling} with $N_\text{R}=2$ yields critical parameters consistent with DOS data.

%%%%%%%%%%%%%%%%%%%%%%%%%%%%%%%%%%%55
\paragraph*{Two terminal conductance:}
%%%%%%%%%%%%%%%%%%%%%%%%%%%%%%%%%%%55
The disorder averaged two-terminal conductance $\overline{g}$ along the z-direction of the sample with length $L$ and width $W$ is obtained from the Landauer formulation of transport, i.e., $g=(e^2/h) \text{Tr} (t^\dg t)$, where $t$ is the transmission matrix. 
We employ the quantum transport code {\tt Kwant}~\cite{Groth2014Kwant}: two infinitely long cubic leads are attached in the z-direction. 
%
%\rim{what does the following half-sentence mean?} 
The lead is represented by the same Hamiltonian~\eqref{eq:Ham0} as in the scattering region with $h=0$. 

Figure~\ref{fig:scaling}(c) shows  $\overline{\ln g}$, which acts as an order parameter~\cite{AbrahamsPRL79, SlevinPRL01} at the metal-insulator transition, for several combinations of $L, W$, and $h$. 
As it turns out, within this setup, the conductance value is small $g\ll 1$ at the critical point, which is related to the choice of the parameter $h=0$ in the lead. 
This, however, does not affect the scaling analysis. 
On the insulating side,  $\overline{\ln g}$ decreases as expected from $g \sim \exp(-L/\xi)$. In contrast, in the Hopf insulating side, there is an initial increase of  $\overline{\ln g}$, as the localization length $\xi \gtrsim L$, and with increasing $L$, the data crosses over to insulating behavior. 
Plotting as a function of scaling variable $(h -\hc)/\hc \cdot L^{1/\nu}$ in Fig.~\ref{fig:scaling}(e), the metallic branch is visible on the Hopf-side, and only when $L>\xi$, it shows the scaling of the insulating phase. 
The scaling exponent $\nu\simeq 1.02(4)$ is consistent with all the previous observables.

%%%%%%%%%%%%%%%%%%%%%%%%%%%%%%%%%%%55
\paragraph*{Dynamical scaling:}
%%%%%%%%%%%%%%%%%%%%%%%%%%%%%%%%%%%55
We calculate the DOS using the standard Kernel polynomial method (KPM)~\cite{KPMRMP}. Here, the DOS is expanded in terms of Chebyshev polynomials, $T_n(x)$, 
$
\rho(\varepsilon) = \text{Tr} \delta (\varepsilon-H) \approx \mu_0 + 2 \sum_{n =1}^{N_\text{m}} g_n \mu_n T_n(\varepsilon). 
$
The trace, $\text{Tr}$, is stochastically estimated using $N_\text{tr}=8$ random vectors in a $128^3$ system. We compute Chebyshev moments, $\mu_n$, and apply the Jackson kernel, $g_n$, to mitigate the Gibbs oscillations arising from a finite number of terms in the Chebyshev expansion $N_\text{m}$.

\begin{figure}[!t]
    \centering
    \includegraphics[width=0.9\columnwidth]{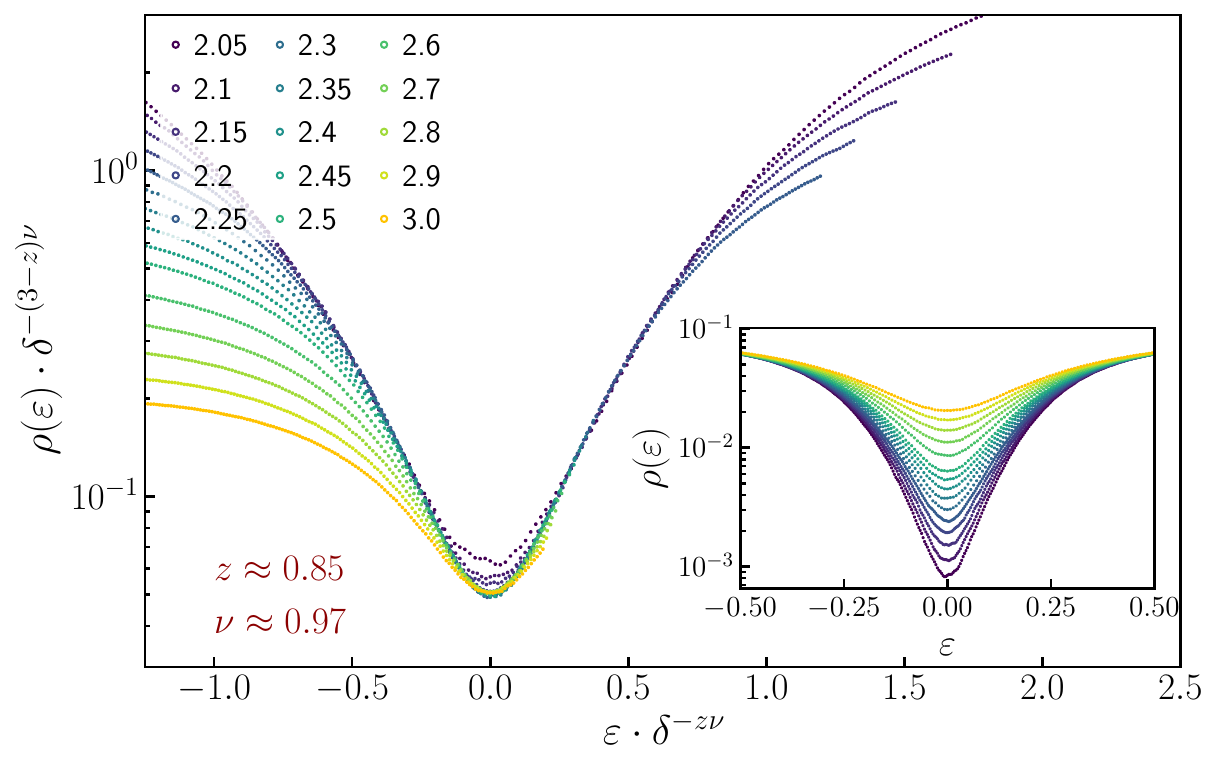}
        \caption{Scaling collapse of the density of states at finite energy $\varepsilon$ for $L=128$ on the metallic side for different disorder values mentioned in the legend. Inset shows the raw $\rho(\varepsilon)$ before rescaling. At energies $\varepsilon \approx 0$, we observe the effects of the finite number of expansion moments, $N_m{=}8193$. We average over approximately $100$ disorder configurations.} 
    \label{fig:sdos}
\end{figure}

\begin{figure}[!t]
    \centering
    \includegraphics[width=0.9\columnwidth]{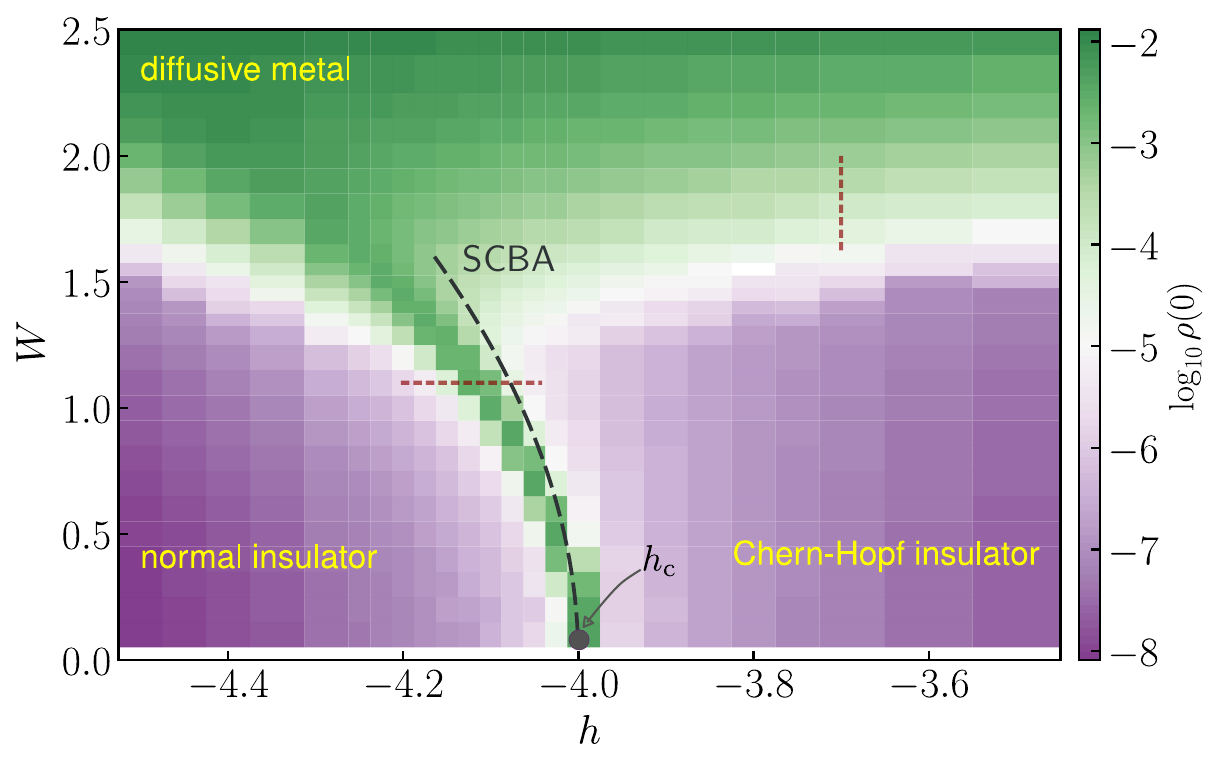}
        \caption{Location of insulating regions, as diagnosed by the bulk density of states at zero energy. 
        %, $\rho(\varepsilon=0)< 10^{-4}$. 
        %Bulk density of states at zero energy, $\log_{10} %\rho(\varepsilon=0)$, calculated using the Kernel %polynomial method. 
        The observed finite width between two insulating phases arises from the limited resolution in the density of states calculation within the KPM method ($N_m{=}4097$ moments and $8$ random vectors are used), with $h_\text{c}$ marking the gap closing in the clean model~\eqref{eq:Ham0}. The dashed horizontal and vertical lines highlight the analysis points for finite-size scaling performed in Figs.~\ref{fig:scaling}, \ref{fig:sdos}. The black dashed line shows the SCBA calculation of the phase boundary. }
    \label{fig:phasediag}
\end{figure}

At finite energy, we assume the following scaling form of the density of states close to the transition approaching from the metallic side, 
$$
\rho(\varepsilon) \sim \delta^{(d-z)\nu} \msr{F}(|\varepsilon| \delta ^{-z \nu}),
$$where $d=3$, and $z$ is the dynamical critical exponent, and $\delta=(W-W_\text{c})/W_\text{c}$. 
$\msr{F}$ is an unknown scaling function that depends on the energy $\varepsilon$, and we expand it up to the second order in the scaling variable.  
Figure~\ref{fig:sdos} shows the rescaled $\rho(\varepsilon) \cdot \delta^{-(3-z)\nu} $ against the scaling variable $\varepsilon \delta^{-z \nu}$ for several values of disorder $W$ for fixed $h=-3.7$ as shown with a vertical dashed line in the  Fig.~\ref{fig:phasediag}, i.e., unlike previously, now approaching the transition from the metallic side with fixed $h$. 
The inset shows the raw DOS data for those disorder strengths. 
The single parameter scaling yields a dynamical exponent as $z\simeq 0.85(5)$. 
This could be contrasted with the following observation that 
%which is consistent \rim{this is not obviously consistent?} %with the observation that 
near the quantum critical point, at small energies  $\rho(\varepsilon)\sim \varepsilon^2$, suggesting $z\approx 1.0$ from the scaling analysis~\cite{HerbutPRL14, LiuPRL16, BeraPRB16,pixley_PRB2016}. This can be attributed to the Dirac nature of the low-energy dispersion at the clean bulk gap closing point $h=-4t$. At the $\Gamma$ point The dispersion can be derived from the continuum description $E(k_x, k_y) \simeq \pm \Le[4t_2^2 (k_x^2+k_y^2)\Ri]^{1/2}$, which is independent of the $k_z$.

%%%%%%%%%%%%%%%%%%%%
\paragraph*{Phase diagram:}
%%%%%%%%%%%%%%%%%%%%
%
Figure~\ref{fig:phasediag} shows the phase diagram of the CHI~\eqref{eq:Ham0} in the presence of disorder close to a gap closing point, $h=-4.0$. 
Due to finite $N_\text{m}$, the $\rho(0)$ is always finite even in the insulating phase. % 
Using $\rho(0) \lesssim 10^{-4}$  as an indicator of the vanishing density of states, we observe an increase of the Chern-Hopf phase with increasing $W$, indicated by the re-entrant phase boundary similar to the disordered topological phase in 2D~\cite{ProdanPRL10, MateoAP23}. 
At higher disorder values%, the $\rho(0) \gtrsim 10^{-4}$,  
the system enters the diffusive metal phase. 
One would expect the standard Anderson localization transition to occur at even higher disorder values, which we do not probe in this study.

%%%%%%%%%%%%%%%%%%%%
%%%%%%%%%%%%%%%%%%%%
%%%%%%%%%%%%%%%%%%%%
\paragraph*{Self-consistent Born Approximation (SCBA):}
%%%%%%%%%%%%%%%%%%%%
 In the presence of disorder, the Green’s function of the electron obeys
$G(\omega,{\bm k})=(\omega-H({\bm k})-\Sigma)^{-1}$. Here, the self-energy of the electron can be evaluated within the self-consistent Born approximation~(SCBA): 
\begin{align}
\Sigma(\omega)=\frac{W^2}{3}\int \frac{d^3k}{(2\pi)^3} (\omega-H({\bm k})-\Sigma+i0^{+})^{-1}.
\label{eq:self_energy} 
\end{align}
Eq.~(\ref{eq:self_energy}) can further be written as\cite{groth_PRL2009}
\begin{align}
\Sigma=\Sigma_0 \mathbb{I}_2+\Sigma_z\sigma_z,
\end{align}
where $\mathbb{I}_2$ is a $2\times 2$ identity matrix, $\Sigma_0=(\Sigma_{11}+\Sigma_{22})/2$, $\Sigma_z=(\Sigma_{11}-\Sigma_{22})/2$. Since the self-energy is momentum
independent for $\delta$-function correlated disorder, it simply
modifies the parameters of the system as 
$\tilde h=h+{\textrm Re}\Sigma_z$, $\tilde E_F= E_F+{\textrm Re}\Sigma_0$,
where `Re' refers to the real part (of the self-energy). Note that $\Sigma_x(y)$ vanishes due to symmetry.

To compute $\Sigma_z$, we set $\Sigma=0$ on the right-hand side of Eq.~(\ref{eq:self_energy}). This leads to 
\begin{align}
\Sigma_z=\frac{W^2}{3} \int \frac{d^3k}{(2\pi)^3} \frac{d_z}{E_F^2-|d({\bm k})|^2}\ .
\end{align}
Setting $E_F=0$, we solve for $\Sigma_z$ numerically at the gapless point with $h=-4$. Interestingly, we find $\Sigma_z>0$. Consequently, the gapless point $h=-4$ shifts to $h-\Sigma_z$, enhancing the topological phase boundary as shown in Fig.~(\ref{fig:phasediag}).

%%%%%%%%%%%%%%%%%%%%
%%%%%%%%%%%%%%%%%%%%

%%%%%%%%%%%%%%%%%%%%
\section{Conclusion and Outlook}
%%%%%%%%%%%%%%%%%%%%
%\SB{This will be taken care by me...}
We have studied the stability of the bulk gapped and boundary gapless states in a 3D CHI. Since there is no time-reversal or charge-conjugation symmetry, it is not obvious whether the topology of the system is protected against disorder. 
Interestingly, similar to conventional topological insulators, the disorder can drive a topological transition, and the gapless boundary states remain stable up to moderate disorder values. 
Investigating the Hopf-metal phase transition's critical properties reveals a correlation length exponent of $\nu \simeq 1.0(1)$, deviating from traditional 3D delocalization-localization transitions across Wigner-Dyson universality classes (orthogonal $\nu \approx 1.571$~\cite{RodriguezPRL10, Slevin_2014}, unitary $\nu \approx 1.43$~\cite{SlevinPRL97}, symplectic $\nu \approx 1.375$~\cite{AsadaJSPS05}), rather revealing the underlying Chern universality of the clean topological transition~\cite{Caio2019}.
In comparison, a recent study in 3D orthogonal class with particle-hole symmetry undergoing topological to metallic transitions shows deviation from the standard universality class $\nu \approx 0.8$~\cite{ShindouPRB20}.

In a situation where the transition is between semimetal and metal, such as the  Weyl semimetal to diffusive metal, the correlation length exponent is found to be $\nu\approx 1.0$~\cite{HerbutPRL14, BeraPRB16, BitanPRX18}.  
The dynamical exponent at that transition was $z\approx 3/2$ due to the linear density of states $\rho(\varepsilon)\sim |\varepsilon|$ at the transition. 
In contrast, approaching the critical point from the metallic phase, we probe the dynamical exponent when $\rho(\varepsilon) \sim \varepsilon^2$ and finite at small energies, where one expects it to be $z\approx 1.0$; however, within our numerical accuracy we observe the slightly smaller value of $z\approx 0.85$. 

%One also observes an exonent 
%

We finally turned to Hopf insulators with zero Chern numbers. The phase diagram for these systems (data not shown here) is expected to be similar to that of the Hopf-Chern insulator. While we expect the critical properties to be identical, a detailed investigation is beyond the scope of this work due to computational limitations stemming from the long-range hopping inherent in the pure Hopf model. Moreover, identifying topological invariants in disordered Hopf-Chern insulators remains a challenge.

In closing, we explore possible experimental avenues for realizing such models and their associated physics. Magnetic materials are highlighted as promising candidates for Hopf insulators, although specific materials fitting this description have yet to be identified. Layered Chern insulators with a twist~\cite{Kennedy_PRB2016} present promising options for Hopf insulators with non-zero Chern numbers. Consequently, helical magnets are considered optimal for realizing Hopf-Chern insulators and exploring disorder effects within these systems.

\section{Acknowledgements}
  We would like to thank Bitan Roy and Robert-Jan Slager for a critical reading of the manuscript and for several insightful discussions.  SB acknowledges support from MPG for funding through the Max Planck Partner Group at IIT Bombay. SB and KS thank the MPI-PKS, Dresden computing cluster, and the visitor program for their hospitality during the project. This work was partly supported by the Deutsche Forschungsgemeinschaft under the grant cluster of excellence ct.qmat (EXC 2147, project-id 390858490). KS also acknowledges funding from the Science and Engineering Research Board (SERB) under SERB-MATRICS Grant No. MTR/2023/000743 and support from the Department of Atomic Energy (DAE), Govt. of India, through the Basic Research in Physical and Multidisciplinary Sciences project via RIN4001.

\appendix

\section{Local charge impurities} \label{app:onsite}
%%%
In this section, we probe the stability of the transition and the associated critical properties by changing the nature of the disorder potential. Here we take the following onsite disorder potential  $H_\text{dis} = \sum_{\br }| \br \ra U({\br})\, \sigma_0 \la \br |$. $U(\br)$ is the strength of the potential, which is distributed over the interval $[-W, W]$. 
%%%%%%%%%%%%%%%%%%%%
\begin{figure}[!h]
    \centering
    \includegraphics[width=1\columnwidth]{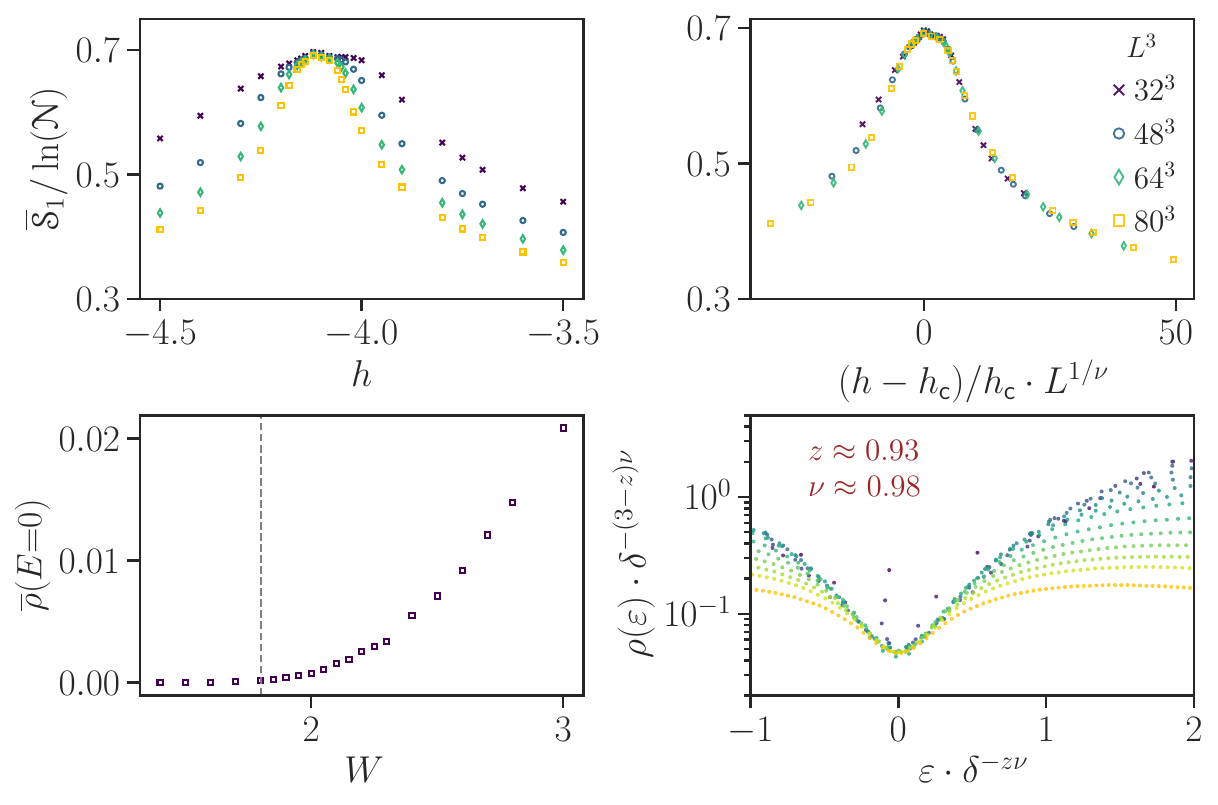}
        \caption{Scaling collapse for the participation entropy $\overline{\msr{S}}_1$ and also the density of states $\rho(E)$. The data supports $\nu\approx 1$, and the dynamical exponent $z\approx 0.93$. } 
    \label{fig:app_scaling}
\end{figure}
%%%%%%%%%%%%%%%%%%%%

The upper panel of Fig.~\ref{fig:app_scaling} shows the scaling of $\overline{\msr{S}}_1/\ln(\msr{N})$  as $h$ is varied for different system sizes of linear length $L=32-80$. The data clearly shows the transition between two different insulating phases at $\hc\simeq -4.115$, and the corresponding scaling collapse is shown with $\nu\approx 1$.  

The lower panel shows the $E=0$ density of states $\overline{\rho}$ calculated using the KPM method described in the main text.   $\rho$ becomes finite with increasing disorder, signifying the transition to the metallic phase. The scaling collapse of the metallic side's finite energy $\rho$ supports critical exponents  $z\approx 0.93$ and $\nu \approx 1$, which agrees with the staggered disorder data shown in the main text.

%%%%%%%%%%%%%%%%%%
\section{Fate of the surface state in the (100)-direction in the presence of disorder} \label{app:xsurface}

%%%%%%%%%%%%%%%%%%%%
\begin{figure}[!h]
    \centering
    \includegraphics[width=1\columnwidth]{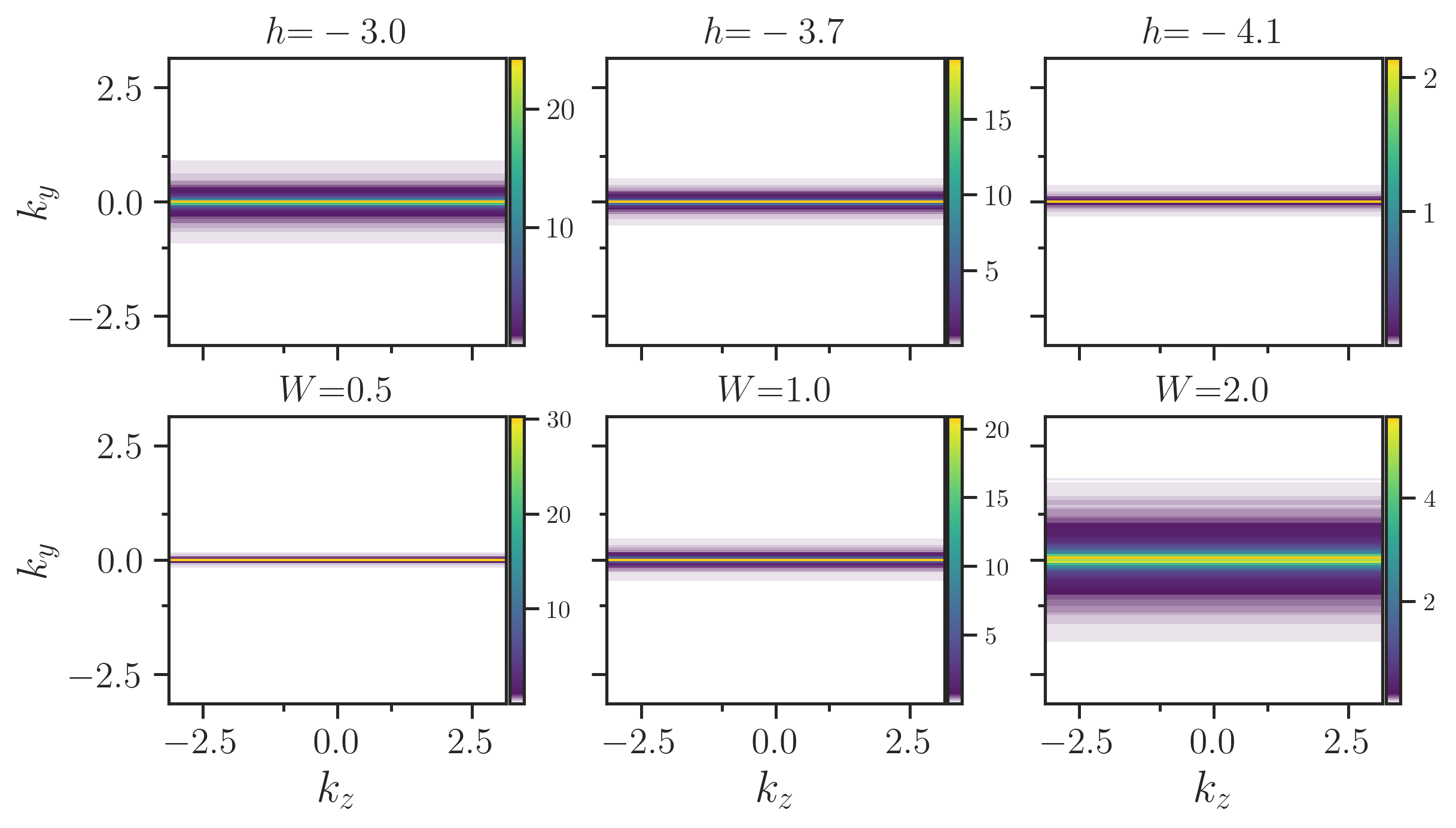}
        \caption{Evolution of the surface spectral function at $E{=}0$ along the (100)-direction. All the parameters are the same as in Fig.~\ref{fig:dis_surface}. System size 
        \{$Lx, Ly, Lz\} = \{64, 128, 128$\}. } 
    \label{fig:xsurface}
\end{figure}
%%%%%%%%%%%%%%%%%%%%

Figure~\ref{fig:xsurface} shows the evolution of the surface spectral function $\msr{A}(k_y, k_z, E{=}0)$~\eqref{eq:spectral} along the (100)-direction, with similar features observed in the (001) direction. At finite disorder $W=1.1$ (upper panel of Fig.~\ref{fig:xsurface}), transitioning from the topologically non-trivial phase to the trivial phase results in a significant reduction in the surface spectral weight, indicating the absence of the state in the trivial phase. Additionally, changing the disorder $W$ at fixed $h=-3.7$, we observe the disappearance of the surface state as the system transitions into the diffusive metal phase, as mentioned in the main text.

%%%%%
\section{Effective Surface Hamiltonian} \label{app:surfaceHam}
\begin{figure}[!t]
\includegraphics[width=6.3cm]{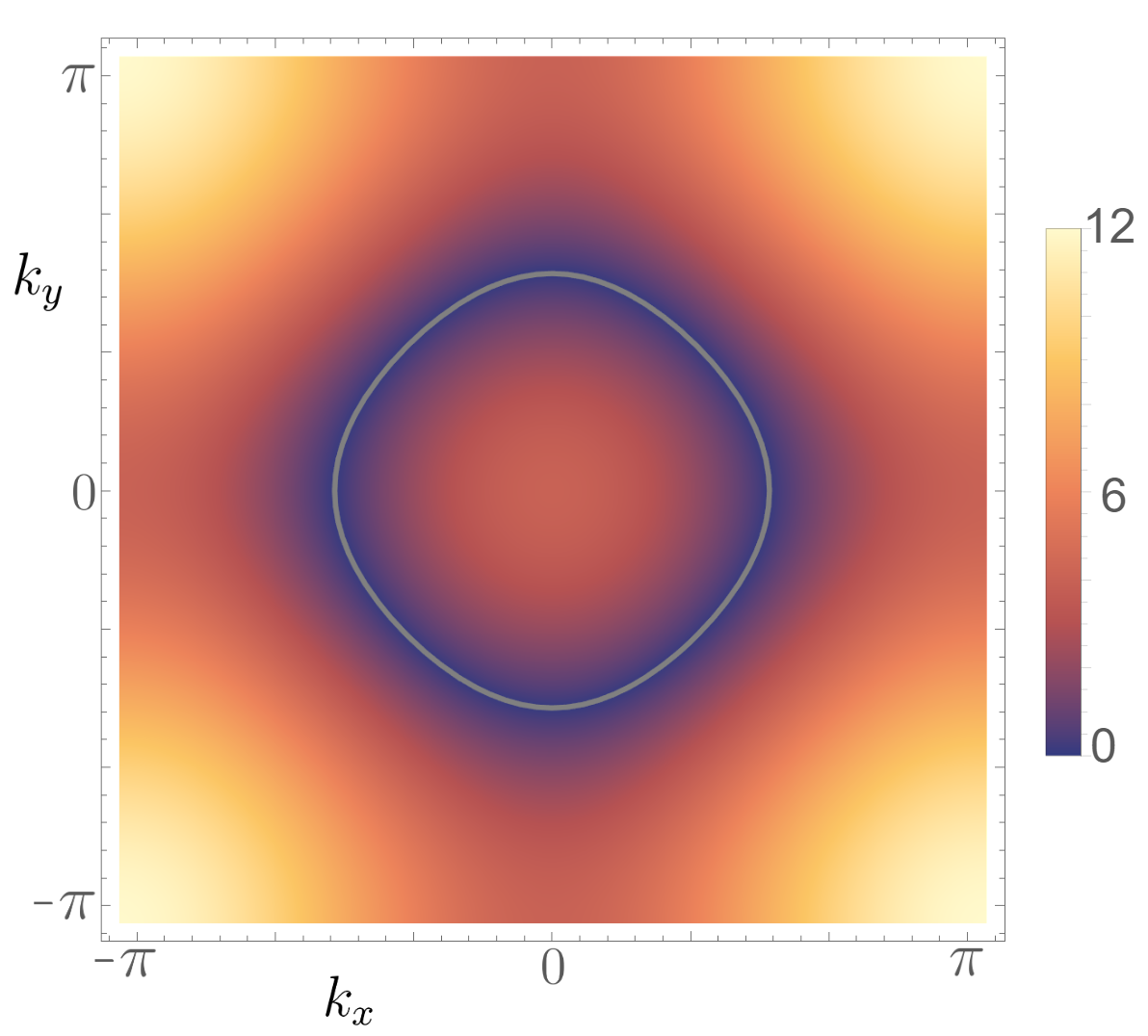}
\caption{
The contour plot illustrates the energy difference $\Delta E=E_c-E_v$ between the two middle bands of the original surface state Hamiltonian, revealing a gapless nodal ring where $\Delta E = 0$. This alignment precisely coincides with the emergence of the gapless ring originating from $d_z$, as depicted by the white line.
}
\label{fig:Effctive_surface}
\end{figure}
%%%%
In this section, we derive the effective Hamiltonian for the surface state along (001) direction. We terminate the lattice in the $\hat{z}$-direction, confining it within $z=1$ and $z=L$, where $L$ is the lattice length. Then, the momentum space Hamiltonian for the Chern-Hopf insulator discussed in the main text can be expressed as
\begin{align}
    \mathcal{H}(k_x, k_y) = \sum_{k_x, k_y, z=1}^{z=L} M_0 (k_x, k_y)\, c_{k_x, k_y, z}^{\dag}\,c_{k_x, k_y, z}\nonumber\\+ \mathcal{T} (k_x, k_y)\, c_{k_x, k_y, z}^{\dag}\,c_{k_x, k_y, z+1} + h.c.\label{eq:surfHam}
\end{align}
where $M_0 (k_x, k_y) = (h + 2 t (\cos(k_x) + \cos(k_y))) \sigma_z$ and $\mathcal{T} (k_x, k_y) = (-i t_2 \sin(k_x) + t_2 \sin(k_y)) \sigma_x + (i t_2 \sin(k_y) + t_2 \sin(k_x))\sigma_y$.

To find the effective surface states, we choose the ansatz $\psi(z)=\lambda^z \phi$ to describe the localized, gapless surface states at the boundary. With this, the boundary equation becomes   
\begin{align}
    d_z\sigma_z\phi + \lambda^{-1} t_2\,(\sin k_y+i\sin k_x) \sigma_x \phi\nonumber \\+ \lambda^{-1} t_2(\sin k_x-i\sin k_y)\sigma_y \phi = E \phi.
\end{align}
Solving this equation, we obtain $E=\pm d_z$,
%This leads to an energy dispersion%\rim{why dispersion? dispersing as a function of what?} of $\pm d_z$,
suggesting that the effective surface state Hamiltonian is $H_{\rm eff}=d_z \sigma_z$.  Fig.~(\ref{fig:Effctive_surface}) demonstrates an exact correspondence between the effective surface state (depicted as a white ring) and the surface state found numerically by solving Eq.~(\ref{eq:surfHam}). 
It turns out that for the other two directions, the effective surface state Hamiltonians are harder to find analytically. Thus, we only show the numerical solutions of the surface states in Fig.~\ref{fig:surfacemodes}.

\bibliography{references}

\end{document}